\documentstyle[12pt]{article}
\title{
\phantom. \hfill{$\scriptstyle {IFUM\ -\ 456/FT}$} \\
\phantom. \hfill{$\scriptstyle {January\ 1994}$}\\
\phantom. \hfill\\
The spatial statistical properties\\
of wave functions in a disordered finite\\
one-dimensional sample.}
\author{{\sl I.V.Kolokolov}
\\ INFN, Sez.di Milano, via Celoria 16,
20133 Milano, Italy and \\ Budker Institute of Nuclear
Physics,\\ Novosibirsk 630090, Russia} \date{}

\begin{document}
\maketitle

\begin{abstract}
For a given wave function one can define a quantity $\mu_E$
having a meaning of its inverse spatial size. The Laplace transform
of the distribution function $P(\mu_E)$ is calculated analytically
for a 1D disordered sample with a finite length $L$.
\end{abstract}

\newpage
\maketitle

\section{Introduction.}
%******************************************************
The Anderson localization was related firstly to the properties
of eigenfunctions of Hamiltonians with random parameters \cite{Anderson},
for review see \cite{Lifshits}. It has been recognized recently that this
phenomenon has more general nature and it is an integral part of
chaotic dynamics of deterministic quantum systems. Initially such an
understanding came into existence from the numerical investigations
\cite{Chir},\cite{Cas},\cite{Izr}.
The ensuing theoretical development \cite{Fe1}
gave rise an idea of the possibility to describe statistical properties
of quantum chaotical systems in terms of 1D and quasi-1D random potential
problems. In this context a detailed knowledge of the wave function
statistics in a disordered sample is of great importance. The case of
finite length $L$ of the sample is of interest too in the light of
numerical experiments \cite{Cas2},\cite{Shep}. On the other hand, the
$L$-dependence is a new degree of freedom and gives new physical
information.

For a given eigenstate $\Psi_E$ one can introduce a quantity $\mu_E$
(see below (\ref{size-def}))
having a meaning of inverse size of the corresponding
wave packet. If some parameters of the Hamiltonian are distributed over
a statistical ensemble the quantity $\mu_E$ becomes a random variabile too.
The distribution function $P(\mu_E)$ reconstructed from a given ensemble
of Hamiltonians plays an important role in interpretation of numerical
experiments \cite{Cas3} and is sensitive to fine features of quantum mechanical
disorder.
This function has been calculated analytically in quasi-one dimensional
case for an infinite sample in the paper \cite{Fe2}.
The method of direct computation of all the moments $<\mu_E^n>$ has been
used. In the same situation the first moment $<\mu_E>$ was found in
\cite{Fe3} for an arbitrary length $L$.

In the present paper I calculate the Laplace transform ${\cal P}(s)$
of the function  $P(\mu_E)$ in strictly 1D case for
an arbitrary value of $L$. The $L\to\infty$ limit coincides with the
quasi-1D result of  \cite{Fe2}. I use essentially the outcomes of the novel
path integral approach to the 1D random potential problem  \cite{Kol-L}.
\footnote {In this paper the first two moments of $P(\mu_E)$
have been found for $L\to\infty$ independently
on \cite{Fe2}.}

The advantages of the method allow to avoid a cumbersome direct
computations of all the moments $<\mu_E^n>$ to restore ${\cal P}(s)$
(or  $P(\mu_E)$).

I consider here one-dimensional Schroedinger problem with the
Hamiltonian:
\begin{equation}\label{Hamiltonian}
\hat{\cal H}=-\frac{d^2}{dx^2}+U(x),
\end{equation}
where $U(x)$ is a random function of $x$ obeying the white-noise
Gaussian statistics:
$<U(x)U(x^\prime)>=D\delta (x-x^\prime)$.
The corresponding statistical ensemble can be described by the measure
of averaging:
\begin{equation}\label{Noise}
{\cal D}U\exp\left(-\frac{1}{2D}\int\limits_{-L}^{L}U^2(x)\,dx\right),
\end{equation}
where $(-L,L)$ is  is the interval, which our system takes up.
For an eigenfunction $\Psi_E$ of
$\hat{\cal H}$:
\[
\hat{\cal H}\Psi_n(x)=E_n\Psi_n(x)
\]
with an eigenvalue $E$ the quantity $\mu_E$ can be introduced as:
\begin{equation}\label{size-def}
\mu_E=\int\limits_{-L}^{L} dx\,|\Psi(x)|^4.
\end{equation}
(It is called also as "inverse participation ratio".)
The "fast-phase limit" $D/E^{3/2}\gg 1$  is proper for the class of
problems mentioned above and it will be used below. Then the random
potential problem (\ref{Hamiltonian}), (\ref{Noise}) is equivalent to
the Abrikosov-Ryzhkin model  (\cite{Ryzh}, for details see  \cite{Kol-L})
solved exactly in  \cite{Kol-L}.
It was obtained there that the multipoint correlator
\begin{eqnarray}\label{p-qm-de}
 & p^{(q,m)}_{E}(x_{1}, x_{2},\dots x_{m-1}, x_{m})=
\nonumber\\
 & =
\langle\sum\limits_n\delta(E-E_n)|\Psi_n(x_{1})|^{2q}
|\Psi_n(x_{2})|^{2q} \dots|\Psi_n(x_{m-1})|^{2q} |\Psi_n(x_{m})|^{2q}\rangle
\end{eqnarray}
is equal to the quantum mechanical matrix element:
\begin{eqnarray}\label{p-qm}
 & p^{(q,m)}_{E}(x_{1}, x_{2},\dots x_{m-1}, x_{m})=
\nonumber\\
 &={\displaystyle \frac{1}{2\pi k (2\alpha)^{qm-1} (qm-2)!}}
\times
\nonumber\\
&\times\langle
e^{\xi/2}|e^{-(L-x_1)\hat{H}}e^{-q\xi}e^{-(x_2-x_1)\hat{H}}
e^{-q\xi} \dots e^{-(x_{m-1}+L)\hat{H}} e^{-q\xi}
e^{-(x_{m}+L)\hat{H}}
|e^{\xi/2}\rangle.
\end{eqnarray}
\[
x_1>x_2>\dots>x_m
\]
with the "effective" Hamiltonian ${\hat H}$ equal to
\begin{equation}\label{Ham-eff}
\hat{H}=-\frac{\alpha}{2}\partial_{\xi}^{2}+
\frac{1}{2\alpha}e^{-\xi}+\frac{\alpha}{8}.
\end{equation}
In the thermodinamic limit $L\to\infty$ we can use the asymptotic
relation:
\begin{equation}\label{Asymps}
\exp(-T \hat{H})\,
e^{\xi/2}\longrightarrow
\Upsilon_{0}(\xi)=\frac{2}{\alpha}
K_{1}\left(\frac{2}{\alpha}e^{-\xi/2}\right)
\end{equation}
following from
\begin{equation}\label{zeromod}
{\hat H}\Upsilon_{0}(\xi)=0.
\end{equation}
(Here $K_{\mu}(z)$ is the standard notation for the modified
Bessel function.)
The explicit expressions for the correlators (\ref{p-qm-de}) are given
in \cite{Kol-L}. For our aims we need only the representation
(\ref{p-qm}).

\section{The distribution function of $\mu_E$ in the infinite
sample limit.}

All the moments $<\mu_E^n>$ of the quantity $\mu_E$ can be expressed
as integrals of correlators $p^{(2,n)}(x_1,\dots, x_n)$ with
respect to the arguments $x_1,..,x_n$. For a given set $<\mu_E^n>$,
$n=0,1,..$ one can restore the Laplace transform ${\cal P}(s)$
of the distribution function $P(\mu_E)$ as:
\begin{equation}\label{Lapl}
{\cal P}(s)=\sum\limits_{n=0}^{\infty} \frac{(-s)^n}{n!}
<\mu_E^n>=
\end{equation}
\[
=\sum\limits_{n=0}^{\infty} (-s)^n \frac{\pi k}{L}
\int\limits_{-L}^{L}\,dx_1
\int\limits_{-L}^{x_1}\,dx_2\dots p^{(2,n)}(x_1,x_2,\dots,x_n).
\]
The factor $\pi k/L$ cancels the high-energy limit of the denstity
of states inherent in the definition of the correlators (\ref{p-qm-de}).

To make the presentation more transparent I start from the
limit $L\to\infty$.
In this case  $p^{(2,n)}(x_1,\dots, x_n)$ is a function of
relative distances only. Thus we can replace $dx_1$--integration
by the factor $2L$ fixing $x_1=0$:
\begin{equation}\label{Pdif}
{\cal P}_{\infty}(s)
=1+2\pi k\sum\limits_{n=0}^{\infty} (-s)^n
\int\limits_{-L}^{0}\,dx_2
\int\limits_{-L}^{x_2}\,dx_3\dots p^{(2,)n}(0,x_2,\dots,x_n).
\end{equation}
In the limit $L\to\infty$ the relation (\ref{p-qm})
with (\ref{Asymps}) leads to the following
matrix element representation:
\begin{eqnarray}\label{p-qm-th}
 & p^{(q,n)}_{E,L\to\infty}(x_{1}, x_{2},\dots x_{n-1}, x_{n})=
{\displaystyle \frac{1}{2\pi k (2\alpha)^{qn-1} (qn-2)!}}
\times
\nonumber\\
&\times\langle
\Upsilon_{0}(\xi)|e^{-q\xi} e^{-(x_1-x_2)\hat{H}}
e^{-q\xi}e^{-(x_2-x_3)\hat{H}}e^{-q\xi}\dots e^{-q\xi}
e^{-(x_{n-1}-x_{n})\hat{H}}e^{-q\xi}
|\Upsilon_{0}(\xi)\rangle.
\end{eqnarray}
Here $x_1>x_2>\dots>x_{n}$. The factorial factor can be transformed
into a power one with the use of
the identity:
\begin{equation}\label{ide}
\frac{1}{(z-1)!}=\frac{e}{2\pi} \int\limits_{-\infty}^{+\infty}
\,d\tau \displaystyle{\frac{e^{i\tau}}{(1+i\tau)^z}}
\end{equation}
Substituting (\ref{p-qm-th}) with this representation for the
$n$-dependent numerical coefficients
into (\ref{Pdif}) and changing the order of summation and
$d\tau$-integration we find that the resulting integrand is just
the expansion of the exponential of the operator
$-2L{\hat Q}$ where
\begin{equation}\label{NovaH}
{\hat Q}={\hat H}-\kappa e^{-2\xi}, \;
\;\;\; \kappa=\frac{-s}{(2\alpha)^2(1+i\tau)^2}
\end{equation}
Thus ${\cal P}_{\infty}(s)$ can be represented in the form:
\begin{equation}\label{P-me}
{\cal P}_{\infty}(s)
=1+\frac{\alpha e}{\pi}
\int\limits_{-\infty}^{+\infty}\,d\tau e^{i\tau}(1+i\tau)\kappa
\langle\Upsilon_{0}|e^{-2\xi}|\Phi(t=0,\xi)\rangle,
\end{equation}
where the function $\Phi(t,\xi)$ is solution of the evolution
equation:
\begin{equation}\label{Evo}
\partial_{t}\Phi=
-{\hat Q}\Phi=-({\hat H}-\kappa e^{-2\xi})\Phi
\end{equation}
with the initial condition $\Phi(t=-L,\xi)=\Upsilon_{0}(\xi)$.
In the $L\to\infty$ limit we need the steady-state solution $\Phi_0$
only:
\begin{equation}\label{Steady}
\Phi\to\Phi_0,\; \;\;\;{\hat Q}\Phi_0=0.
\end{equation}
The initial condition fixes the leading $\xi\to\infty$ asymptotics:
%\begin{equation}\label{Asss}
$\Phi_0(\xi\to\infty)\rightarrow e^{\xi/2}$.
%\end{equation}
The corresponding solution of the eq.(\ref{Steady}) has the form:
\begin{equation}\label{Steady-sol}
\Phi_0(\xi)=e^{\xi/2}\Gamma(1-\lambda)
W_{\lambda,1/2}\left(\sqrt{\frac{-8\kappa}{\alpha}} e^{-\xi}\right).
\end{equation}
Here $W_{\lambda,1/2}$ is the Whittaker function and
$\lambda=-(-8\kappa \alpha^3)^{-1/2}$.
Turning back to the expression (\ref{P-me}) and noting that
$\kappa e^{-2\xi}\Phi_0={\hat H}\Phi_0$ (see (\ref{Steady})
and (\ref{zeromod})) we obtain:
\begin{equation}\label{Pass}
{\cal P}_{\infty}(s)
=1+\frac{\alpha^2 e}{2\pi}
\int\limits_{-\infty}^{+\infty}\,d\tau e^{i\tau}(1+i\tau)
\left(\Upsilon_{0}^{\prime}\Phi_0-\Upsilon_{0}\Phi_0^{\prime}
\right)|_{\xi\to +\infty}.
\end{equation}
Using the expansion of  $W_{\lambda,1/2}(z)$ at $z\to 0$:
\[
W_{\lambda,1/2}(z)\approx \frac{1}{\Gamma(1-\lambda)}\left[
1-\frac{z}{2} + \lambda z \left(\psi(1)+\psi(2)-\psi(1-\lambda)
-\ln z\right)\right]
\]
(see (\cite{Grad});
$\psi(a)$ is the logarithmic derivative of the Gamma function)
and the analogious expansion of $K_1(z)$
we find:
\begin{equation}\label{Puss}
{\cal P}_{\infty}(s)
=\frac{e}{\pi}
\int\limits_{-\infty}^{+\infty}\,d\tau e^{i\tau}(1+i\tau)
\psi\left(1+\frac{2(1+i\tau)}{\sqrt{\alpha s}}\right)=
{\displaystyle
\frac{\alpha s}{2\sinh^2(\sqrt{\alpha s}/2)}}.
\end{equation}
(In taking the integral the property $\psi(z)=\psi(z+1)-1/z$
has been used.)
The surprising thing is that the result (\ref{Puss}) coinsides
with the formula obtained in \cite{Fe2} for the quasi-1D case.
It means that the distribution function ${\cal P}(s)$ is insensitive
in the quasi-1D sample to transverse fluctuations. (The author is
greateful to Y.V.Fedorov for this interpretation.) The asymptotics
of $P(\mu_E)$ at $\mu_E\to 0$ following from  (\ref{Puss})
(see \cite{Fe2}) $P(\mu_E)\sim(\alpha/\mu_E)^{3/2}\exp(-\alpha/2\mu_E)$
corresponds to the zero probability of non-localized states.

\section{The distribution function of $\mu_E$ for finite
samples .}

In this section we obtain an analytical expression for ${\cal P}(s)$
when $\alpha L$ is finite. (The inequality $kL\gg 1$ still holds.)
Substituting the quantum mechanical representation (\ref{p-qm}) into
the general expression (\ref{Lapl}) and using the identity (\ref{ide}) we
have:
\begin{equation}\label{Lapl-L}
{\cal P}(s)=1+
\frac{\alpha e}{2\pi L}
\int\limits_{-\infty}^{+\infty}\,d\tau e^{i\tau}(1+i\tau)
\times
\end{equation}
\[
\times\sum\limits_{n=1}^{\infty} \kappa^n
\int\limits_{-L}^{L}\,dx_1
\int\limits_{-L}^{x_1}\,dx_2\dots
\langle
e^{\xi/2}|e^{-(L-x_1)\hat{H}}e^{-2\xi}e^{-(x_2-x_1)\hat{H}}
\dots e^{-2\xi}
e^{-(x_{n}+L)\hat{H}}
|e^{\xi/2}\rangle,
\]
where $\kappa$ and $\hat{H}$ are defined in equations (\ref{Ham-eff})
and (\ref{NovaH}).
The operator $\hat{\Pi}(t)$:
\begin{equation}\label{Pip}
\hat{\Pi}(t)=
\sum\limits_{n=1}^{\infty} \kappa^n
\int\limits_{-L}^{t}\,dx_1
\int\limits_{-L}^{x_1}\,dx_2\dots
e^{-(t-x_1)\hat{H}}e^{-2\xi}e^{-(x_2-x_1)\hat{H}}
\dots e^{-2\xi}
e^{-(x_{n}+L)\hat{H}}
\end{equation}
obeys the inhomogenious differential equation:
\begin{equation}\label{Pipur}
\partial_t\hat{\Pi}=
-{\hat Q}\hat{\Pi}+\kappa e^{-2\xi}
e^{-(t+L)\hat{H}}
\end{equation}
and the initial condition $\hat{\Pi}(t=-L)=0$.
(The operator ${\hat Q}$ is defined in  (\ref{NovaH}).) The
solution of this operator equation has the form:
\begin{equation}\label{Pipsol}
\hat{\Pi}(t)=
\kappa
\int\limits_{-L}^{t}\,dt^{\prime}
e^{-(t-t^{\prime})\hat{Q}}e^{-2\xi}
e^{-(t^{\prime}+L)\hat{H}}
\end{equation}
Thus the function ${\cal P}(s)$ can be expressed as:
\begin{equation}\label{PPpip}
{\cal P}(s)=1+
\frac{\alpha e}{2\pi L}
\int\limits_{-\infty}^{+\infty}\,d\tau e^{i\tau}(1+i\tau)
\langle
e^{\xi/2}|
\hat{\Pi}(L)
|e^{\xi/2}\rangle =
\end{equation}
\[
=1+
\frac{\alpha e}{2\pi L}
\int\limits_{-\infty}^{+\infty}\,d\tau e^{i\tau}(1+i\tau)
\kappa \int\limits_{-L}^{+L}\,dt
\langle
e^{\xi/2}|e^{-(L-t)\hat{Q}}e^{-2\xi}
e^{-(t+L)\hat{H}}
|e^{\xi/2}\rangle.
\]
Let us represent in the latter matrix element the left function
$e^{\xi/2}$ as $e^{\xi/2}=\Phi_0(\xi)+w(\xi)$ and the right one
as $e^{\xi/2}=\Upsilon_0(\xi)+y(\xi)$. The functions $w(\xi)$ and
$y(\xi)$ are normalizable and can be expanded over a complete
set in the functional space. The functions $\Phi_0(\xi)$ and
$\Upsilon_0(\xi)$ are exact zero modes of the operators
${\hat Q}$ and $\hat{H}$ correspondingly.
Thus the term containing $\Phi_0(\xi)$ as bra-vector and
$\Upsilon_0(\xi)$ as ket-vector with the addendum $1$ coincide
with the right hand side of (\ref{Pass}) and reproduce the
result (\ref{Puss}) for ${\cal P}_{\infty}(s)$. Evaluating
the remaining terms using again the equalities
${\hat Q}\Phi_0=\hat{H}\Upsilon_0=0$
and $e^{-2\xi}=(\hat{H}-{\hat Q})/\kappa$,
\footnote {Representing $e^{-2\xi}$ by means of this identity
we arrive at the total derivative under the sign of $dt$-integration.
The latter, thus, can be performed explicitly already at this stage.}
we obtain:
\begin{equation}\label{pro}
{\cal P}(s)={\cal P}_{\infty}(s)-
\frac{\alpha e}{2\pi L}
\int\limits_{-\infty}^{+\infty}\,d\tau e^{i\tau}(1+i\tau)
\left(\langle w|w\rangle -\langle w| \exp(-2L{\hat Q})|w\rangle
\right).
\end{equation}
The $\tau$-independent terms are omitted in the integrand in
eq. (\ref{pro}) since they give zero result after
$d\tau$-integration.
One can calculate analytically the contribution to
${\cal P}(s)$ of the first term in square brackets using the
integral representation of the Whittaker function \cite{Grad}
and performing first the $d\tau$-integration.
The final result has the form:
\begin{equation}\label{P}
{\cal P}(s)=
{\cal P}_{\infty}(s)+\frac{1}{L}F_1(s)+\frac{1}{L}
{\displaystyle e^{-\alpha L/4}} F_2(s,L),
\end{equation}
where ${\cal P}_{\infty}(s)$ is the limit of the distribution
function when $L\to\infty$ (see (\ref{Puss})),
the function $F_1(s)$ does not depend on $L$:
\begin{equation}\label{P-1/L}
F_1(s)=-{\displaystyle
\frac{s}{2\sinh^2(\sqrt{\alpha s}/2)}}
\ln \left(1+\sinh^2(\sqrt{\alpha s}/2)\right),
\end{equation}
and $F_2(s,L)$ goes to 0 when $L\to\infty$:
\begin{equation}\label{P-exp}
F_2(s,L)=\frac{s}{\pi} \int\limits_{-\infty}^{+\infty}\,d\tau
{\displaystyle e^{-i\tau\sqrt{2\alpha s}}}
\int\limits_{0}^{\infty}
{\displaystyle \frac{d\nu 2\nu\sinh 2\pi\nu}{\cosh\pi (\nu-\tau)}
e^{-\alpha L\nu^2}}
V(\tau,\nu)\,V(\tau,-\nu).
\end{equation}
Here
\begin{equation}\label{Vv}
V(\tau,\nu)=\int\limits_{0}^{\infty}\,\frac{du}{u}
W_{i\tau,i\nu}(u)\left(\Gamma(1-i\tau)
W_{i\tau,1/2}(u)-1\right)
\end{equation}
and $W_{a,b}$ is the standard notation for the Whittaker function
\cite{Grad}.
The power-like tail is given by the second term in (\ref{P}) and
for $\alpha L/4\geq 1$ the function ${\cal P}(s)$ can be approximated
with a good precision by first two terms in (\ref{P}).

\section{Acknowledgments}

I am very grateful to
Yan V. Fyodorov for fuitful discussions and
stimulating advices.
I would like to express my thanks to Profs. G.Casati and
B.V.Chirikov for useful conversations.

%************************************************************
%****************************************************************

%%%%%%%%%%%%%%%%%%%%%%%%%%%%


\newpage

\begin{thebibliography}{99}

\bibitem{Anderson} Anderson P. Phys. Rev. 109 (1958),1492.

\bibitem{Lifshits}Lifshits I.M., Gredeskul S.A, Pastur L.A
Introduction to the theory of disordered systems.
John Wiley and Sons, Inc., 1988.

\bibitem{Chir}Chirikov B.V. Phys. Rept. 52 (1979), 263.


\bibitem{Cas}Casati G., Chirikov B.V.,
Guarnery I., Shepelyansky D., Phys. Rept. 154 (1987), 77.

\bibitem{Izr}Izrailev F.M., Phys. Rept. 196 (1990), 299.

\bibitem{Fe1}Fyodorov Y.V., Mirlin A.D., Phys.Rev.Lett.,67(1991), 2405.

\bibitem{Cas2}Casati G., Molinari L., Israilev F.M.,
Phys.Rev.Lett. 64(1990), 1851.

\bibitem{Shep}Shepelyansky D.L.  Physica D, 28 (1987), 103.

\bibitem{Cas3}Casati G., Chirikov B.V.,
Guarnery I., Izrailev F.M., Phys. Rev.E 48 (1993), R1613.

\bibitem{Fe2}Fyodorov Y.V., Mirlin A.D., Phys.Rev.Lett.,
71(1993),412;\\
Mirlin A.D, Fyodorov Y.V., J.Phys.A 26(1993), L551.

\bibitem{Fe3}Fyodorov Y.V., Mirlin A.D., Phys.Rev.Lett.,69(1992),1093.

\bibitem{Kol-L}Kolokolov I.V., JETP 76(1993), 1099.

\bibitem{Ryzh}Abrikosov A.A., Ryzhkin I.A. Adv.in Phys. 27 (1978),
146

\bibitem{Grad} Gradstein I.S., Ryzhyk I.M., Tables of Integrals,
Series and Products. Academic (N.Y.), 1980.


\end{thebibliography}
\end{document}